# Microstructure, morphology and lifetime of armored bubbles exposed to surfactants**

*Anand Bala Subramaniam, Cecile Mejean, Manouk Abkarian, Howard A. Stone**

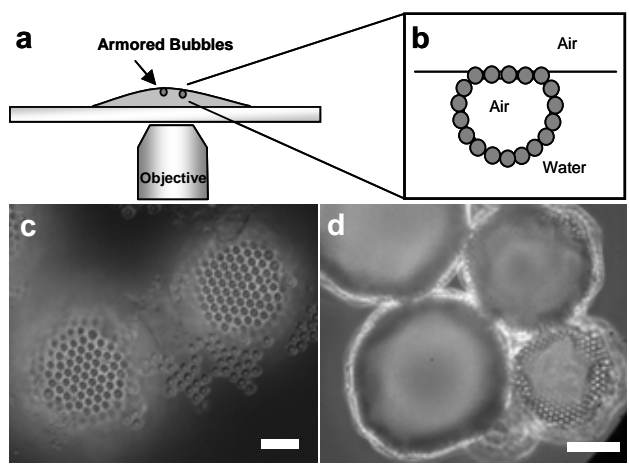

**We report the behavior of particle-stabilized bubbles (armored bubbles) when exposed to various classes and concentrations of surfactants. The bubbles are non-spherical due to the jamming of the particles on the interface and are stable to dissolution prior to the addition of surfactant. We find that the dissolving bubbles exhibit distinct morphological, microstructural, and lifetime changes, which correlate with the concentration of surfactant employed. For low concentrations of surfactant an armored bubble remains non-spherical while dissolving, while for concentrations close to and above the surfactant CMC a bubble reverts to a spherical shape before dissolving. We propose a microstructural interpretation, supported by our experimental observations of particle dynamics on the bubble interface, that recognizes the role of interfacial jamming and stresses in particle stabilization and surfactant-mediated destabilization of armored bubbles.**

It is well known that surfactants stabilize foams and emulsions as well as individual bubbles [1, 2]. Also, it has been long recognized that foams and emulsions can be stabilized by colloidal particles [3, 4], though detailed studies of the particle-enhanced stabilization have only been performed in recent years [5]. In particular, some cases have been reported where individual bubbles, protected by a close-packed shell of particles (colloidal armor), appear to be stable indefinitely and do not undergo the usual disproportionation [6, 8].

Nevertheless, many systems contain both particles and surfactants and a wide variety of responses have been reported [5, 9]. Alargova et. al. report that the addition of sodium dodecyl sulfate (SDS), an anionic surfactant, significantly reduced the stability of a foam stabilized by rod-shaped SU-8 particles (results were reported at one concentration, 10 wt% SDS, where the primary mode of destabilization was gravity induced film drainage and foam collapse) [7]. They attribute their observations to a change in the surface properties of the particles, from hydrophobic to hydrophilic, due to the oriented adsorption of SDS onto the particles. This change decreases the particles' affinity for the interface and consequently the foam has a lifetime comparable to that of an SDS-stabilized foam [7].

Here we show that the behavior of particle-stabilized bubbles in the presence of surfactant is richer than what is suggested by Alargova et al's original observations. Indeed, our observations of individual particle-covered bubbles in the absence of surfactant reveal that they adopt stable non-spherical shapes. These non-spherical bubbles dissolve when exposed to surfactant and exhibit distinct morphological changes that are dependent on the concentration of surfactant employed. Direct microstructural observations of the particle shell leads us to propose that wetting changes are only sufficient to explain destabilization at relatively high surfactant concentrations, while the interfacial stresses on the shell that accompany bubble dissolution plays an important role in destabilizing the bubbles for lower surfactant concentrations. Our results provide insights into the still open question of the mechanism of particle super-stabilization of bubbles in the absence of surfactants and shows promise as a novel means of controllably releasing colloidal particles and gases.

In our experiments, we synthesized monodisperse particle-covered bubbles (armored bubbles) through a microfluidic hydrodynamic focusing method [10] or by simply shaking an aqueous suspension of particles to obtain polydisperse bubbles (see Experimental Section). Our quantitative experiments were performed with a model suspension of 4.0 μm diameter surfactant-free charge-stabilized polystyrene particles and the non-ionic surfactant Triton X- 100. We also conducted qualitative experiments with a variety of particles and surfactant types to obtain a general overview of the surfactant destabilization phenomenon (Supplementary Information).

We place a surfactant-free aqueous sample containing armored bubbles onto a glass slide and we observe that the buoyant armored bubbles rise to the top of the droplet (Figure 1a). In this configuration, the shell of particles deforms at the bulk air-water interface to produce a distinct flat facet (Figure 1b,c). Our observations indicate that the particles form a bridge between the air phase in the bubble and the atmosphere, with only a thin layer of water in between. Such a bridging effect was shown previously in a

*Figure 1:* a) Schematic of the experimental setup. A drop of sample containing the armored bubbles is placed on a slide and visualized from below using an inverted microscope. b) The buoyant bubbles rise to the surface of the drop and deform to form a facet at the interface. c) A top view of the armored bubbles taken with an upright microscope. The particles composing the facet are in fact bridging the two air phases, that in the bubble and that in the atmosphere, with an intervening film of water. Scale bar 16 μm. d) Viewed from below: the bubbles progressively lose gas and take on buckled non-spherical shapes before stabilizing. The ability to sustain non-spherical shapes, as shown here, is a hallmark of solid-like behavior. Scale bar 48 μm.

* A. Bala Subramaniam, C. Mejean, Dr. M. Abkarian, Professor H. A. Stone
  Division of Engineering and Applied Sciences
  Harvard University
  Pierce Hall, 28 Oxford Street, Cambridge, MA 02138 (USA).
  Fax: (617) 495-9837
  E-mail: has@deas.harvard.edu

** We thank the Harvard MRSEC and Unilever for support. We also thank D. Gregory and colleagues for helpful conversations.

Supporting information for this article is available on the WWW under http://www.angewandte.org or from the author.

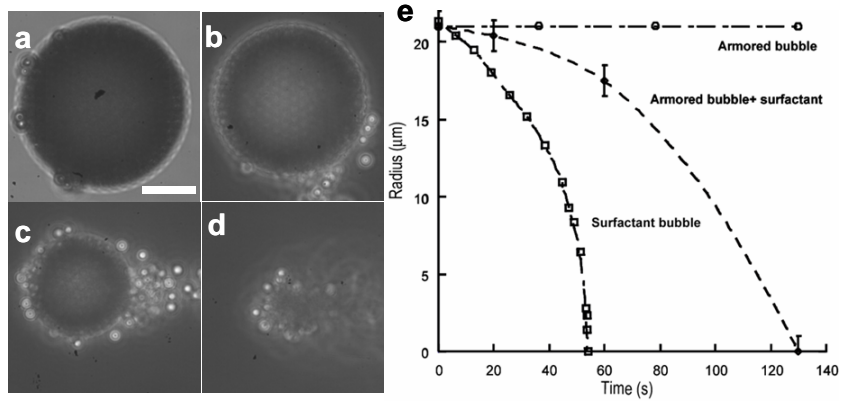

**Figure 2:** Response of armored bubbles to 0.66 mM Triton X-100 (surfactant); bubbles were covered with polystyrene particles. Scale bar 24 μm. a) The initially stable non-spherical bubble regains its spherical shape by rapidly ejecting excess particles from the interface. b,c) The bubble continues to shed particles smoothly as it decreases in size until d) it disappears. e) Radius versus time for a stable armored bubble, an armored bubble exposed to surfactant, and a surfactant-covered bubble under the same experimental conditions. Note that in the presence of surfactant the particles do not halt dissolution, but slow it down relative to that of a bare surfactant-covered bubble.

system of particle-covered oil droplets [11,12]. In the case of gas bubbles, the bridging effect we demonstrate is significant since it allows for the rapid diffusion of gas out of the bubble directly into the atmosphere, through the thin water film.

Viewing through the water phase (i.e. with an inverted microscope) we observe that the bubble progressively takes on a non-spherical shape by losing some gas, after which the non-spherical bubble then remains stable without further changes in volume or shape for at least two days (Figure 1d). The observation of buckling and the stable non-spherical shapes suggest that the interface of the armored bubble is behaving in a solid-like manner, since equilibrium non-spherical shapes of ordinary bubbles with an isotropic surface tension is prohibited. This solid-like behavior was recently proposed to arise from the jamming of the adsorbed particles on the bubble interface [10,13,14] and is to be contrasted with the apparent maintenance of sphericity in other experiments [6].

We next systematically investigate the influence of surfactants on the structure and stability of the armoured bubbles. When a solution of Triton X-100 at a concentration, c=0.66 mM (critical micellar concentration, CMC approximately 0.2 mM), was added to the sample, the non-spherical bubble quickly regains a spherical shape (within one camera frame, or one second) by ejecting excess particles from the interface (Figure 2a). The remaining particles on the interface resume Brownian motion in a well-defined hexagonally coordinated lattice (See Supplementary Movie 1 where the dynamics of the particles are clearly visible). The spherical bubble now proceeds to dissolve continuously until it disappears completely (Figure 2a-d, Supplementary Movie 2). This observation is reproducible and, generally, armored bubbles with an initial radius of 20 μm takes 100 ± 3 s to dissolve when exposed to this concentration of surfactant. The response is to be contrasted with the apparent infinite lifetime of an armored bubble in an air-saturated solution, and the comparatively shorter lifetime of 50 s for a simple surfactant-covered bubble (Figure 2e). CMC (0.066 mM), markedly different behavior is observed. The bubble remains non-spherical but starts ejecting particles while losing volume (Figure 3a-d). Moreover, the particles remain immobile on the interface, signifying their still jammed state (Supplementary Movie 3). The bubble continues to lose gas, with periods of transient stability when the bubble does not change in apparent radius (here the apparent radius is taken to be one half the diagonal length of the smallest bounding rectangle). Moreover, unlike the case for c>CMC, the lifetime of individual bubbles are highly variable, ranging from 1190 seconds to 1340 seconds for bubbles with an initial radius of 20 μm (Figure 3e). Nevertheless, despite the differences in response of individual bubbles, all the bubbles eventually dissolve completely.

We constructed a phase diagram of bubble behavior by exposing individual bubbles to various concentrations of surfactant (Figure 4). The boundary $c_{critical}^{(2)}$ in the phase diagram, which denotes a change from an aspherical, unstable, jammed shell to a spherical, unstable shell with evident Brownian motion, is not sharply delineated due to the complicated adsorption pattern in these systems. The surfactant adsorbs on the particles and the air-water interface [15,16], both of which change in area as the bubble dissolves and particles are ejected, thus preventing the accurate determination of adsorbed amounts and its correspondence to the surfactant CMC. However, we note that this critical concentration is close to the CMC of the surfactants tested, where surfactant adsorption onto surfaces is expected to be maximal [15].

We now seek to rationalize the clear difference in bubble lifetime, microstructure and morphology, which correlates with surfactant concentration. For $c > c_{critical}^{(2)}$ the microstructural signature is the unjamming of the shell and the morphological signature is the rapid return to a spherical shape. Consistent with Alargova et. al's observations for a different system [7], we believe that at this concentration range the surfactant acts as a detergent, promoting the wetting of the particles by adsorbing onto the particles and the air-water interface. Since confinement on an interface is a requirement for interfacial jamming[10], desorption should lead to the unjamming of the shell. Nevertheless, we note that the morphological and microstructural changes we observe suggest that particle desorption is rapid (< 1s) yet the particles still remain in a spherical shell-like configuration for the duration of the bubble lifetime (Figure 2). A plausible explanation for this surprising observation is that the surfactant-covered particles, despite being completely wetted by the liquid, experience an attraction to the surfactant-covered air-water interface, and thus do not diffuse away. The origin of this long-range attraction remains to be elucidated, though there have been reports of long-range attraction between surfactant-covered surfaces in other systems [17].

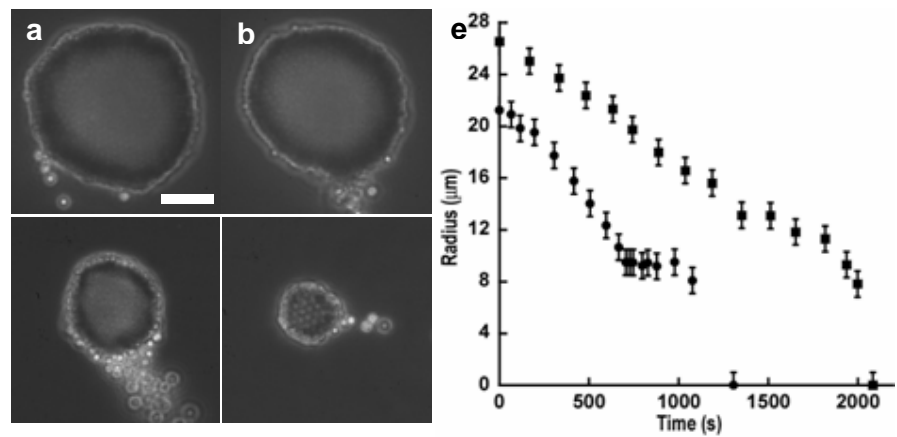

**Figure 3:** Response of armored bubbles to 0.066 mM Triton X-100. Scale bar 24 μm a, b) The bubble maintains a non-spherical shape but is no longer stable and starts shrinking by losing particles (bright white circles). c,d) The non-spherical shape is maintained throughout, until the bubble disappears. e) Plot of apparent radius (half the diagonal length of the smallest bounding rectangle) versus time of two different armored bubbles exposed to the same experimental conditions. Note that these bubbles takes much longer to dissolve compared to those shown in Figure 2 and exhibits transient plateaus where the apparent radius does not change. There is also significant variability in the dissolution time of individual bubbles under similar experimental conditions.

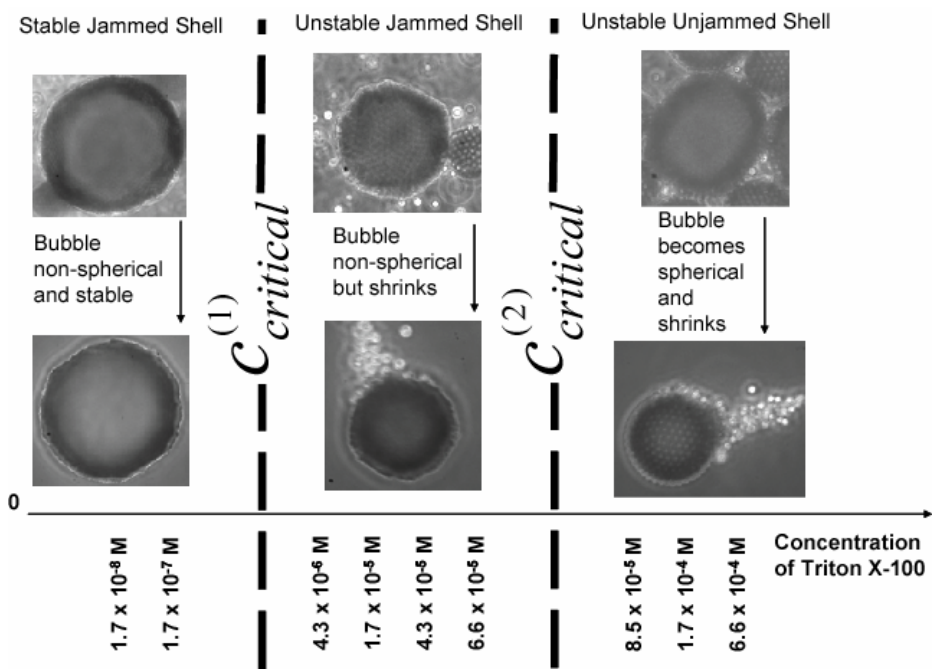

**Figure 4:** Phase diagram of armored bubble response to surfactant. $c_{critical}^{(1)}$ signifies the critical surfactant concentration above which the armored bubbles are no longer stable to dissolution, while $c_{critical}^{(2)}$ denotes the boundary of the unjamming transition where the particles on the shell resume Brownian motion and the bubble reverts to a sphere. Lifetime of the bubble decreases as the surfactant concentration increases.

The scenario in which the particles are desorbed and only remain *close* to the interface also accounts for the smooth dissolution profile and the shorter lifetime of the bubbles in the regime of c> $c_{critical}^{(2)}$ (Figure 2e): The free air-water interface is unable to resist the dissolution of the bubble and the particles are released when available surfactant-covered interface decreases as the bubble disappears. Although the particles are not straddling the interface and do not affect the mechanical properties of the interface, the gas impermeable particles still shield the interface and reduce gas flux out of the dissolving bubble, which is a plausible explanation for the longer lifetime of the armored bubbles exposed to surfactants relative to that of a simple surfactant-covered bubble (Figure 2e).

We next consider the observations for c< $c_{critical}^{(2)}$ (Figure 3). It is well known that decreasing bulk surfactant concentration results in both smaller changes in wetting properties and surface tension, and also an increase in the amount of time required for adsorption [1, 15]. Thus, for a small enough surfactant concentration, no destabilization should occur which is consistent with our observations for c< $c_{critical}^{(1)}$, which is the lowest concentration range reported in the phase diagram (Figure 4). A possible trivial explanation for the longer timescale for destabilization in the intermediate concentration range, $c_{critical}^{(1)}$ <c< $c_{critical}^{(2)}$ may be that the surfactant takes a longer time to adsorb onto the particles, and cause particle desorption. However, changes in wetting properties do not account for the microstructural and morphological observations of the non-spherical shell (Figure 3, 4). Moreover, given the significant variability in lifetime of bubbles exposed to similar experimental conditions (Figure 3e), we hypothesize that the jamming and the consequent stress-bearing capabilities of the shell might play a role.

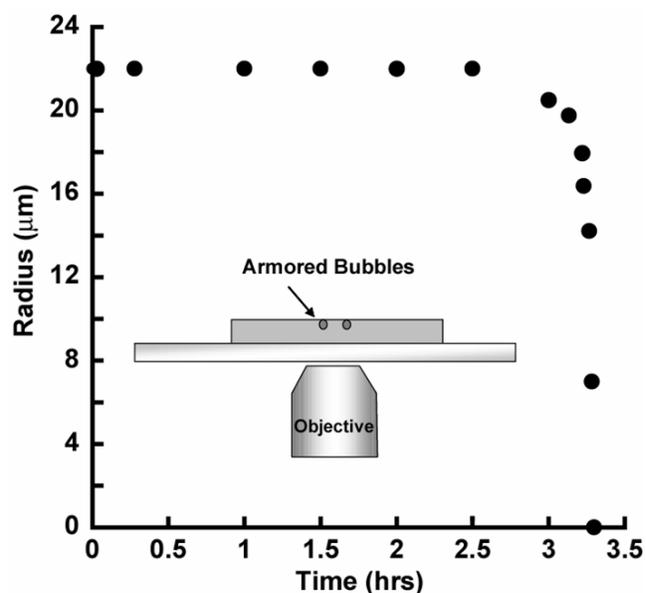

**Figure 5:** Inset, schematic of the experimental setup. The bubbles were exposed to 0.066 mM Triton X-100, similar to the concentration employed for the bubbles in Figure 3. Here, the bubbles are kept from direct contact with the atmosphere (compare to Figure 1) thus, the gas in the bubble has to diffuse through the bulk fluid phase towards the chamber access ports to equilibrate with the atmosphere. The armoured bubble remains stable for about 3 hours before beginning to dissolve. No particles are ejected prior to the start of the dissolution process. Note that time for surfactant diffusion and adsorption should be similar as to that in the system open to the atmosphere.

A test of our hypothesis requires the untangling of two apparently coupled timescales, the time for surfactant to adsorb onto the particles and the time for the bubble to shrink. We thus modified the rate of bubble dissolution by placing the bubbles in a semi-closed system, a perfusion chamber with two access ports that were left open to the atmosphere (Figure 5). Unlike the experiments performed above, the chamber shields the armored bubbles from direct contact with the atmosphere. The shielding results in a concentration gradient between gas in the bubble and the atmosphere [18]. This causes gas to diffuse from the bubble through the bulk water phase towards the chamber access ports which are at atmospheric pressure. We find that the lifetime of the bubble when exposed to 0.066 mM Triton X-100 is greatly increased to hours from the mere minutes in the open system (Figure 5). During this time, particles desorbed only when the bubbles reduce in size and not otherwise. Since these experiments modify the rate of dissolution and not the time required for surfactant adsorption, it is now clear that the process of gas dissolution is necessary for the ejection of particles from the bubble interface when $c_{critical}^{(1)} < c < c_{critical}^{(2)}$.

These observations reveal that the particle shell is providing active rather than passive resistance to the dissolution of the bubble. To illustrate the distinction, consider a balloon composed of a thin elastic membrane: inflating the balloon causes the buildup of tensions in the membrane which must resist the internal pressure of the trapped gas. Thus, a blown up spherical balloon is under higher stress than a deflated one. In contrast, in the case of armored bubbles, the deflation of the bubble due to the dissolution process actually increases the stresses on the shell up to the point where dissolution is halted. A two-dimensional theoretical model by Kam and Rossen of armored bubbles suggests that the particles might be bearing significant stress that is distributed homogenously on a spherical shell as the bubble stabilizes [19]. The distribution of the compressive forces is symmetric in their model, akin to a compression dome, and the consequent rigidity of the shell was suggested to be sufficient to support zero or negative capillary pressures [19]. Note however, that our observation of bubble buckling as gas dissolution occurs suggests that the stress distribution on the bubble surface is not homogeneous, i.e. the shell supports local tensile stresses as well as compressive stresses.

Nevertheless, it appears that the addition of surfactant renders the system unstable and stressed particles, which otherwise were stably held in the shell, are now ejected to relieve the stress buildup as the bubble shrinks. The inhomogeneities in stress distribution on the jammed shell and the history dependence of particle contact forces for individual bubbles also provides a rational explanation for the non-monotonic dissolution profile and the varying lifetimes for bubbles when $c_{critical}^{(1)} < c < c_{critical}^{(2)}$ (Figure 3e).

In conclusion, we have shown that isolated particle-stabilized bubbles take on non-spherical shapes as they stabilize. We obtain a general phase diagram of armored bubble behavior versus surfactant concentration and demonstrate that the jammed shell is stressed, a fact which is revealed when the bubble is exposed to surfactant concentrations roughly below the CMC. Since it is apparent that changes in particle wetting are not sufficient to explain particle ejection for $c_{critical}^{(1)} < c < c_{critical}^{(2)}$, the details of surfactant action must be further studied. For concentrations of surfactant close to the CMC, the shell unjams and the bubble rapidly reverts to a spherical shape. Simple detergency is adequate to explain the lifetime and morphological changes of the bubble in this regime, though the source of the apparent long-range attraction between the bubble and the particles remains to be investigated. The observations we report here have fundamental significance for designing better foam and bubble-stabilizing technologies.

## Experimental Section

We employed monodisperse charged-stabilized polystyrene latex particles of diameter 4.0 µm, 1.6 µm diameter silica particles (Bangs Lab), 1.0 µm diameter PMMA particles (Bangs Lab), and polydisperse agglomerated gold microparticles with mean diameters ranging from 1.0–3.99 µm (Sigma). The surfactants used were Triton X-100, sodium dodecyl sulfate (SDS), Tween 20, cetyl triammonium bromide (CTAB), octyl-B-6-glucopyranoside, and Brij 35 all purchased from Sigma. The CMC of Triton X-100 ranges from 0.22-0.5 mM (data from manufacturer). The surfactant solutions were prepared with ultrapure water (Millipore) at concentration ranging from 0.1 to 10 times the CMC.

10 ml of an aqueous suspension of particles at a volume fraction of 0.1 was shaken manually and vigorously, for about 10 s in a 50 ml test tube. The result was a dilute suspension of gas bubbles, each coated with a jammed shell of particles.

For the experiments open to the atmosphere, we deposited 2.5 µl of the bubble solution onto a glass slide. 100 µl of particle-free surfactant solution was deposited near the 2.5 µl sample solution. The large difference in volumes was chosen to obtain a homogenous concentration of surfactant in the entire sample.

We followed the size of the bubbles by capturing pictures with a high-resolution camera. The bubbles were observed using a phase-contrast objective which enhances the contrast of the ejected particles. The bubble diameters were measured within an error of ± 1 µm.

We followed at least three bubbles and repeated the experiment four times for each concentration of surfactant tested. Care was taken to ensure that each system only contained a small number of armored bubbles. Furthermore, the bubble of interest was always isolated from other bubbles by more than two times the bubble diameter in order to avoid multi-body effects[20]. Since the system was exposed to the atmosphere, the concentration of air dissolved in the bulk around each bubble was rapidly equilibrated with atmospheric pressure.

For the experiments in the closed chamber, a Molecular Probes press to seal perfusion chamber (Molecular Probes) was adhered to a glass slide. The chamber had two access ports at either ends through which the reagents were introduced. We placed 2.5 µl of the bubble solution into the chamber, and flushed the chamber with a surfactant solution of known concentration. This ensures that the bubbles are surrounded with liquid of uniform surfactant concentration. The access ports were left open to the atmosphere, and thus gas from the bubble had to diffuse through the bulk fluid phase to equilibrate with atmospheric pressure[18].

All experiments were carried out at room temperature.

*Colloids, Bubbles and Surfactants*

A. Bala Subramaniam, C. Mejean, M. Abkarian, H.A. Stone*

Microstructure, morphology and lifetime, of armored bubbles exposed to surfactants

**Colloidal Comets:** Surfactants can destroy otherwise stable particle-covered bubbles. One such bubble when exposed to a surfactant, dissolves, leaving a comet-like trail of colloidal particles (see picture).

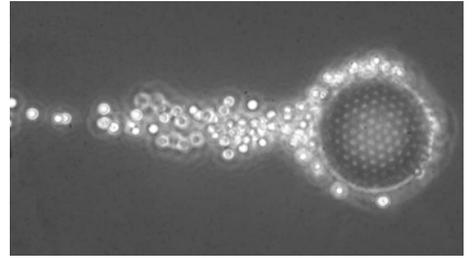

**Caption for Supplementary Movies:**

**Supplementary Movie 1:** Video of the unjammed thermally equilibrated shell after exposure to Triton X-100 at a concentration of 0.66 mM, c> $c_{critical}^{(2)}$. The bubble is covered with 4.0 μm charge stabilized polystyrene beads.

**Supplementary Movie 2:** Time lapse video of an armored bubble dissolving after exposure to Triton X-100 at a concentration of 0.66 mM, c> $c_{critical}^{(2)}$. Note the spherical shape of the dissolving bubble, and the mobility of the particles on the interface. The bubble is covered with 4.0 μm charge stabilized polystyrene beads.

**Supplementary Movie 3:** Time lapse video of an armored bubble dissolving after exposure to Triton X-100 at a concentration of 0.066 mM $c_{critical}^{(1)}$<c<$c_{critical}^{(2)}$. Note the non-spherical shape, and the absence of mobility of the particles on the interface, a signature of a jammed shell.

**Supplementary Information:**

**Reproducibility data for armored bubbles exposed to Triton X-100 at a concentration of 0.66 mM:**

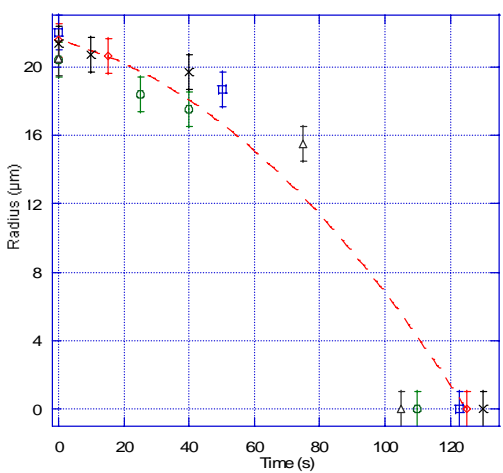

**Supplementary Figure 1:** Radius versus time plots of four bubbles exposed to 0.66 mM Triton X-100 showing the distribution of the time of dissolution of the bubbles.

**Exposure of the armoured bubbles to different surfactant species:**

The armoured bubbles were exposed to an anionic surfactant, sodium dodecyl sulfate, to determine the effect of surfactant type on the destabilization process. The process of destabilization is very similar to the one described for Triton X-100 (Figure 2e in the main paper).

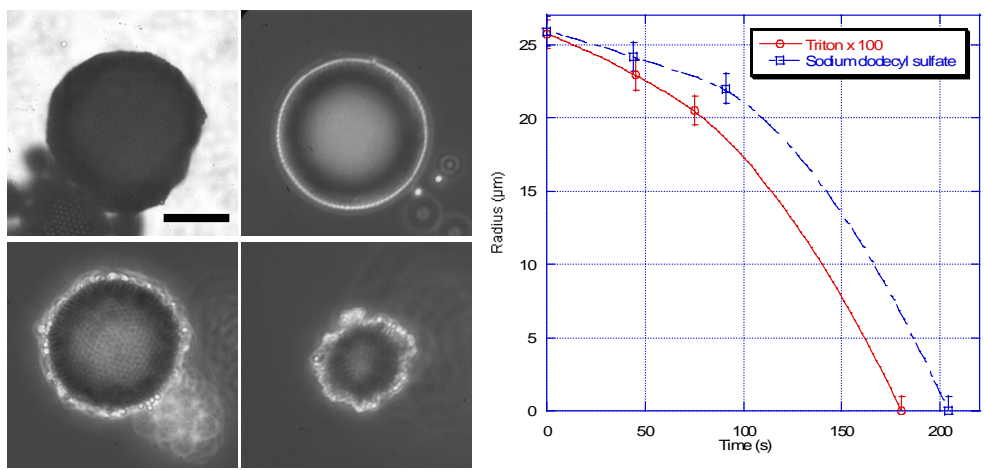

**Supplementary Figure 2:** Exposure of armored bubbles to the anionic surfactant sodium dodecyl sulphate (SDS) at 3 times its critical micellar concentration (CMC). Scale bar 40 μm **a,b)** Note that the initially non-spherical bubble quickly returns to sphericity by losing particles and the proceeds to shrink continuously while ejecting particles, a behavior reminiscent of the non-ionic surfactants. **c,d)** However, unlike exposure to non-ionic surfactants, the particles ejected from the bubble are not colloidally stable, i.e. they are aggregated to each other and often remain associated with the armour shell. The aggregation is simply due to the ions screening the charges on the particles. **e)** Despite the difference in appearance there is no appreciable variation in the bubble lifetime when compared to that obtained after exposure to Triton X-100 at a concentration above its CMC.